\documentclass[journal=jctcce,manuscript=article]{achemso}

\usepackage{graphicx}
\usepackage[multiple]{footmisc}
\usepackage{amsmath}
\usepackage{amssymb}
\usepackage{algorithm}
\usepackage{algpseudocode}
\usepackage{bm}
\usepackage{dcolumn}
\newcolumntype{x}[1]{D{.}{.}{#1}}
\usepackage{txfonts}
\usepackage[rightcaption]{sidecap}
\usepackage{color}
\usepackage[usenames,dvipsnames]{xcolor}
\definecolor{myblue}{rgb}{0,0,1}
\usepackage[breaklinks=true,colorlinks=true,linkcolor=myblue,urlcolor=myblue,citecolor=myblue]{hyperref}
\newcommand{\vk}{{\bm{k}}}
\newcommand{\vG}{{\bm{G}}}
\usepackage{setspace}
\usepackage{enumitem}

\title{An Efficient scaled opposite-spin MP2 method for periodic systems}
\author{Idan Haritan}
\affiliation{The Alexander Kofkin Faculty of Engineering, Bar-Ilan University, Ramat Gan 52900, Israel}
\author{Xiao Wang}
\email{xwang431@ucsc.edu}
\affiliation{Department of Chemistry and Biochemistry, University of California Santa Cruz, Santa Cruz, CA 95064, United States}
\author{Tamar Goldzak}
\email{tamar.goldzak@biu.ac.il}
\affiliation{The Alexander Kofkin Faculty of Engineering, Bar-Ilan University, Ramat Gan 52900, Israel}

\date{Jan 2025}

\begin{document}

\maketitle
\begin{abstract}
We develop SOS-RILT-MP2, an efficient Gaussian-based periodic scaled opposite-spin second-order Møller-Plesset perturbation theory (SOS-MP2) algorithm that utilizes the resolution-of-the-identity approximation (RI) combined with the Laplace transform technique (LT). In our previous work [J. Chem. Phys. 157, 174112 (2022)], we showed that SOS-MP2 yields better predictions of the lattice constant, bulk modulus, and cohesive energy of 12 simple semiconductors and insulators compared to conventional MP2 and some of the leading density functionals. In this work, we present an efficient SOS-MP2 algorithm that has a scaling of $O(N^4)$ with the number of atoms $N$ in the unit cell and a reduced scaling with the number of k-points in the Brillouin zone. We implemented and tested our algorithm on both molecular and solid-state systems, confirming the predicted scaling behavior by systematically increasing the number of atoms, the size of the basis set, and the density of k-point sampling. Using the benzene molecular crystal as a case study, we demonstrated that SOS-RILT-MP2 achieves significantly improved efficiency compared to conventional MP2. This efficient algorithm can be used in the future to study complex materials with large unit cells as well as defect structures.
\end{abstract}

\section{Introduction}
The electronic structure of extended systems, such as solids and other periodic structures, is known to be a computationally complex task due to the effectively infinite number of atoms.
Ab initio quantum chemistry methods for molecules have a long history of developing hierarchical and systematically improvable wavefunction-based methods. Starting from the mean-field Hartree-Fock (HF) approximation, a range of methods are available, such as M{\o}ller-Plesset perturbation theory (MP) and coupled cluster (CC) methods.~\cite{shavitt_many-body_2009,bartlett_coupled-cluster_2007} However, their development and application in the materials science and condensed matter physics communities have been limited, primarily due to the computational challenges posed by extended systems. Instead, density functional theory (DFT)~\cite{hohenberg1964inhomogeneous, kohn1965self} has been the cornerstone of electronic structure calculations for extended systems. DFT is exact in principle and operates on electron density rather than many-body wavefunction, making it a low-scaling theory compared to wavefunction-based methods. However, in practice, using the commonly employed semilocal or hybrid functionals~\cite{becke1993new} can lead to systematic errors in calculating certain properties of materials, such as band gaps, van der Waals interactions, and spectroscopic properties.~\cite{cohen_challenges_2012, perdew_density_1985}  
In recent years, wavefunction methods, such as the periodic extensions of MP and CC, have been successfully applied to condensed phase materials~\cite{Hirata2004,Grueneis2011,Booth2013,Grueneis2015,McClain2017,Tsatsoulis2017,gruber2018,gruber2018a,zhang2019,Gao2020,Nusspickel2022,maschio2007fast,pisani2008periodic,mullan2022reaction}. Nevertheless, the use of these methods is still limited, and further development and applications to periodic systems, especially for complex systems with large unit cells, such surface reactions, van der Waals 2D materials, and point defects, are needed.   
Among these methods, second-order M{\o}ller-Plesset perturbation theory (MP2) is the simplest wavefunction method that captures electron correlation effects, with a computational scaling of $O(N^5)$, where $N$ denotes the number of atoms in the system~\cite{moller1934note}.
As a way to improve MP2 accuracy without additional cost, spin-component-scaled MP2 (SCS-MP2) was introduced by Grimme.~\cite{scs_mp2_grimme2003} In SCS-MP2, the same spin (SS) and opposite spin (OS) contributions to the MP2 correlation energy are separately scaled, leading to the expression for the SCS-MP2 correlation energy:
\begin{equation}
\label{eq:scs-mp2}
    E_\mathrm{SCS-MP2}^\mathrm{corr} = 
    c_\mathrm{os} E_\mathrm{os}^\mathrm{corr}
    + c_\mathrm{ss} E_\mathrm{ss}^\mathrm{corr}
\end{equation}
For $c_\mathrm{os}=c_\mathrm{ss}=1$ one retains the MP2 correlation energy. Building on the SCS concept, various SCS variants of ab initio methods have been developed and tested.~\cite{sos_mp2_head_gordon,hyla2004comprehensive,takatani2007performance, distasio2007optimized,kossmann2010correlated,grimme2012spin,grimme2005accurate,szabados_theoretical_2006,grimme2004calculation,takatani2008improvement,hellweg2008benchmarking}
Among them is the scaled opposite-spin MP2 (SOS-MP2)~\cite{sos_mp2_head_gordon} that retains and
scales only the opposite spin component of the correlation energy, offering further simplifications and computational advantages relative to SCS-MP2. The SCS/SOS variants of MP2 have been shown to outperform conventional MP2 for many molecular properties.~\cite{scs_mp2_grimme2003,hyla2004comprehensive,takatani2007performance,distasio2007optimized,kossmann2010correlated,grimme2012spin} For example, SCS-MP2/SOS-MP2 was found to provide a mean absolute deviation (MAD) of 1.18/1.36 kcal/mol for heats of formation in the G2/97 set of molecules, respectively, significantly better than MP2 (MAD of 1.77 kcal/mol) and the most popular DFT-B3LYP (MAD of 2.12 kcal/mol)~\cite{grimme2005accurate}.
The scaling of OS and SS contributions can be motivated in several ways, including a derivation based on modified perturbation theory~\cite{szabados_theoretical_2006,grimme2012spin}. 
Based on empirical optimization against CCSD(T) reaction energies, the optimal parameters of SS and OS for thermochemical molecular properties proposed first by Grimme are $(c_\mathrm{os}, c_\mathrm{ss}) = (1.2,0.33)$~\cite{scs_mp2_grimme2003}.  

In addition to its comparably good performance~\cite{sos_mp2_head_gordon,grimme2012spin}, SOS-MP2 can be performed with a reduced $O(N^4)$ scaling using the resolution of the identity approximation (RI, also known as density fitting, DF) and the Laplace transform (LT) of the energy denominator~\cite{weigend1997ri,haser1992laplace} (see Section \ref{methods} for details).
For example, Distasio and Head-Gordon~\cite{distasio2007optimized} obtained accurate inter-molecular binding energies for the S22 dataset \cite{jurevcka2006benchmark} 
with the $O(N^4)$ SOS(MI)-MP2 method with reoptimized OS parameters.
Although SCS/SOS methods have been extensively studied in molecular systems, their application and development to real solids and nanostructures are still scarce. In our recent work, we showed that the results of our periodic SCS-MP2 calculations were accurate for the thermochemistry properties of a set of 12 solids, with errors that are smaller than those of the leading density functionals \cite{goldzak2022accurate}. Another recent work by Liang, Ye, and Berkelbach used SCS-MP2 to predict the cohesive energies of molecular crystals.~\cite{liang2023can} It showed that by reoptimizing the SCS parameters, one can achieve $7.5$ kJ/mol accuracy for the cohesive energies in the $X23$ molecular crystal data set.~\cite{dolgonos2019revised} 

This work presents a novel and efficient periodic SOS-MP2 algorithm based on the RI and LT techniques, inspired by algorithms that were proposed for molecules \cite{sos_mp2_head_gordon}, using atom-centered Gaussian orbitals adapted for periodic systems. This algorithm, termed SOS-RILT-MP2, reduces the scaling of conventional MP2 to $O(N^4)$ with the number of atoms per unit cell and to $O(N_k^2)$ the number of k-points sampled in the Brillouin zone. Previously, combining RI and LT was introduced in periodic (conventional) MP2 with atomic orbitals, but has not reduced the computational scaling.~\cite{izmaylov2008resolution,ayala2001atomic, shang2020implementation} On the other hand, quartic scaling MP2 algorithms that incorporate RI and LT have been developed for periodic systems, but were based on plane-wave basis sets~\cite{del2013electron,schafer2017quartic}. Using atom-centered basis sets can offer various advantages over plane waves, such as allowing for straightforward all-electron calculations, easy access to core states, and better convergence behaviors with respect to the basis set size.~\cite{villard2023plane}
 
The structure of this paper is as follows, in Section \ref{methods} we will introduce the SOS-RILT-MP2 formalism and algorithm. Section \ref{results} presents the results of the accuracy and timing of the SOS-RILT-MP2 algorithm in the molecular case for linear alkane chains with increasing length and for two ionic semiconductors, and an example for the benzene molecular crystal. Lastly, we present our conclusions in Section \ref{conc}.

\section{Methods and algorithm }
\label{methods}
We used the single-particle basis of crystalline Gaussian-based atomic orbitals (AOs). These are linear combinations of atom-centered Gaussian orbitals adapted to the translational symmetry of the crystal \cite{McClain2017}.
With periodic boundary conditions and $N_k$ crystal momenta $\vk$ sampled from the Brillouin zone, the MP2 correlation energy can be decomposed into its spin components as follows:
\begin{subequations}
\label{mp2}
\begin{align}
E_\mathrm{os}^{\mathrm{corr}} &= - \frac{1}{N_k^3} \sum_{\vk_i\vk_a\vk_j\vk_b}^\prime \sum_{iajb}
    T_{i \vk_i,j \vk_j}^{a\vk_a, b\vk_b}
    (i \vk_i a \vk_a | j \vk_j b \vk_b),\label{eq:mp2_os} \\
\begin{split}
E_\mathrm{ss}^{\mathrm{corr}} &= - \frac{1}{N_k^3} \sum_{\vk_i\vk_a\vk_j\vk_b}^\prime 
    \sum_{iajb} 
    \left[ T_{i \vk_i,j \vk_j}^{a \vk_a, b\vk_b} 
        - T_{i \vk_i,j \vk_j}^{b \vk_b, a\vk_a} \right]\times (i\vk_i a\vk_a | j\vk_j b\vk_b),
        \label{eq:mp2_ss}
\end{split}
\end{align}
\end{subequations}
where
\begin{equation}
 T_{i \vk_i,j\vk_j}^{a\vk_a,b\vk_b}
    = \frac{(i \vk_i a \vk_a|j\vk_j b\vk_b)^*} { \varepsilon_{a\vk_a} + \varepsilon_{b\vk_b} - \varepsilon_{i\vk_i} - \varepsilon_{j\vk_j} }.
\label{T_amp}
\end{equation}
Electron repulsion integrals (ERIs) are expressed in Mulliken notation $(11|22)$. Throughout this paper, we use $i,j$ to refer to occupied orbitals and $a,b$ virtual orbitals obtained from periodic Hartree-Fock (HF), which we assume to be spin-restricted. The primed summation indicates conservation of crystal momentum, $\vk_a+\vk_b-\vk_i-\vk_j=\vG$, where $\vG$ is a reciprocal lattice
vector. In this work, Gaussian density fitting (GDF), in other words, resolution of the identity (IR) with Gaussian-based auxiliary basis sets, was used to evaluate ERIs~\cite{Sun2017a}. The SCS-MP2 correlation energy will be given by Eq.\ref{eq:scs-mp2}. For SOS-MP2 the same-spin scaling coefficient is $c_\mathrm{ss}=0$, and the optimal opposite spin parameter was empirically found to be $c_\mathrm{os}=1.3$.~\cite{sos_mp2_head_gordon} In our recent work, we showed that this optimal parameter is also valid for calculating the properties of the 12 semiconductors and insulators that we tested.~\cite{goldzak2022accurate}. 

In this work, we develop an efficient algorithm for periodic SOS-MP2 that combines the RI and LT techniques, which we name SOS-RILT-MP2.
First, we will describe the LT algorithm for the denominator of the orbital energy differences in Eq.\ref{T_amp}. We can replace the $1/x$ function using the LT with an infinite integral $\frac{1}{x}=\int_{0}^{\infty} e^{-xt} \,dt $.
In Laplace transformed MP2, the energy denominators are transformed as
\begin{equation}
\frac{1}{ \varepsilon_{a\vk_a} + \varepsilon_{b\vk_b} - \varepsilon_{i\vk_i} - \varepsilon_{j\vk_j}}=\int_{0}^{\infty} e^{-( \varepsilon_{a\vk_a} + \varepsilon_{b\vk_b} - \varepsilon_{i\vk_i} - \varepsilon_{j\vk_j})t}dt
\end{equation}
This Laplace integral is then discretized into a weighted sum of quadrature points.
There are many different algorithms to evaluate the optimal quadrature points for the numerical integral evaluation.~\cite{haser1992laplace,ayala2001atomic,takatsuka2008minimax} 
Alml\"{o}f and Haser were the first to propose an efficient scheme that involved directly minimizing the sum of squares error of the quadrature in order to choose the optimal quadrature points~\cite{almlof1991elimination,haser1992laplace}, and obtained a precision of micro-Hartree ($\mu$H) with a small number of quadrature points. The integral boundaries for each molecule were chosen on the basis of its orbital energy differences. Another algorithm proposed by Ayala, Kudin, and Scuseria introduced a logarithmic transform of the Laplace integration variable~\cite{ayala2001atomic}.
An advancement proposed by Takatsuka, Ten-no, and Hackbusch is to minimize the Chebyshev norm of the quadrature error using the minimax algorithm~\cite{takatsuka2008minimax}. We tried all of the above methods for a set of five different molecules from the S22 set~\cite{jurevcka2006benchmark}, as well as diamond as a representative solid-state system; see the SI for a numerical comparison of all the methods. We found that Takatsuka, Ten-no, and Hackbusch's algorithm was the most efficient among the three methods, having an accuracy of $\mu$H with no more than $6$ quadrature points, and we chose it for our periodic development of Laplace transformed SOS-MP2 (together with RI). 

We will elaborate here on the approximation proposed by Takatsuka, Ten-no, and Hackbusch~\cite{takatsuka2008minimax}. The Laplace transform of a reciprocal $1/x$ in a certain interval $x\in[1,R]$ can be evaluated using the following numerical quadrature:
\begin{equation}
\label{num_int}
\frac{1}{x}\approx E_k(x;{w_l},{t_l})=\sum_{l=1}^{L} w_le^{-xt_l},
\end{equation}
with the roots ${t_l}$ and the weights ${w_l}$. Multiplying the equation by $1/A$ will give the corresponding approximation in an arbitrary interval $y=Ax \in [A,AR]$, such that $\frac{1}{y}\approx E_k(y;{\tilde{w_l}},{\tilde{t_l}})$,
where $\tilde{w_l}=w_l/A, \tilde{t_l}=t_l/A $.  The minimax approximation is used to find the optimal parameters that minimize the Chebyshev norm for each particular $R$. The parameter $R$ is the intrinsic range of the problem and not limited to molecular systems. 
In this work, Laplace decomposition is used in concert with RI, so we will use it for each pair of occupied and virtual energy difference, such that:
\begin{equation}
\label{num_int2}
\int_{0}^{\infty} e^{-(\varepsilon_{a\vk_a}- \varepsilon_{i\vk_i}) t} e^{-(\varepsilon_{b\vk_b}-\varepsilon_{j\vk_j}) t} dt= \sum_l \sqrt{w_l}e^{-(\varepsilon_{a\vk_a}- \varepsilon_{i\vk_i}) t_l} \sqrt{w_l}e^{-(\varepsilon_{b\vk_b}-\varepsilon_{j\vk_j}) t_l}
\end{equation}
The interval will be for each pair of orbitals, such that $\varepsilon_{a\vk_a}- \varepsilon_{i\vk_i} \in [E_{min}, E_{max}]$, and $E_{min}=\varepsilon_{{L}\vk_{L}}- \varepsilon_{{H}\vk_{H}}, E_{max}=\varepsilon_{{max}\vk_{max}}- \varepsilon_{{min}\vk_{min}}$, where $\varepsilon_{{L}\vk_{L}},\varepsilon_{{H}\vk_{H}} $ are orbital energies of the lowest virtual orbital and the highest occupied orbital across all k-points, respectively, and $\varepsilon_{{max}\vk_{max}}, \varepsilon_{{min}\vk_{min}}$ are the maximum and minimum orbital energies across all k-points, respectively. The quadrature points and their weights in the minimax approximation $w_l, t_l $ are determined for $R=E_{max}/E_{min}$. The Laplace-transformed MP2 energies are calculated using $\tilde{w_l}=w_l/E_{min}, \tilde{t_l}=t_l/E_{min}$ (i.e., $A = E_{min}$). The convergence of the quadrature is exponential regardless of the intrinsic range $R$ , resulting in a low number of grid points with a certain accuracy for all problems.
The quadrature points developed by these authors have been kindly made available to the community and a copy can be found on GitLab~\cite{hackbusch2019computation,EXP_SUM_1_x}.

As mentioned above, in this work we will incorporate RI together with LT for the opposite-spin part of the MP2 correlation energy (or simply, SOS-RILT-MP2) in Eq.\ref{eq:mp2_os}. We generalized the SOS-MP2 algorithm developed for molecules to periodic systems~\cite{sos_mp2_head_gordon, takatsuka2008minimax}. 
The RI approximation is used to calculate the ERIs using GDF~\cite{Sun2017a}. We will denote the auxiliary basis functions by $P, Q$ so that we can write the ERIs as:

\begin{equation}
\label{2eri_ri}
(i \vk_i a \vk_a | j \vk_j b \vk_b)=\sum_{P}^{N_{aux}} B_{i \vk_i a \vk_a}^P B_{j \vk_j b \vk_b}^P,
\end{equation}
where 
\begin{equation}
\label{eq:ri_tensor}
B_{i \vk_i a \vk_a}^{P}=\sum_{Q}^{n_{aux}} (i \vk_i a \vk_a |Q)(P |Q )^{-1/2}.
\end{equation}
Thus, the opposite-spin component of the MP2 correlation energy given in Eq.\ref{eq:mp2_os} is expressed utilizing the RI and LT approximations as:
\begin{align}
\label{eq:E_os_RILT}
E^{\mathrm{corr}}_\mathrm{os}& =  - \frac{1}{N_k^3} \sum_l \sum_{\vk_i\vk_a\vk_j\vk_b}^\prime \sum_{iajb}
    (i \vk_i a \vk_a|j\vk_j b\vk_b)^* (i \vk_i a \vk_a | j \vk_j b \vk_b) w_l e^{-(\varepsilon_{a\vk_a}- \varepsilon_{i\vk_i}) t_l} e^{-(\varepsilon_{b\vk_b}-\varepsilon_{j\vk_j}) t_l}\\\nonumber
&= - \frac{1}{N_k^3} \sum_l \sum_{\vk_i\vk_a\vk_j\vk_b}^\prime \sum_{iajb} (\sum_{P}^{N_{aux}} B_{i \vk_i a \vk_a}^P B_{j \vk_j b \vk_b}^P)^* (\sum_{Q}^{N_{aux}} B_{i \vk_i a \vk_a}^Q B_{j \vk_j b \vk_b}^Q) w_l e^{-(\varepsilon_{a\vk_a}- \varepsilon_{i\vk_i}) t_l} e^{-(\varepsilon_{b\vk_b}-\varepsilon_{j\vk_j}) t_l} \\ \nonumber
&= - \frac{1}{N_k^3} \sum_l \sum_{\vk_i\vk_a\vk_j\vk_b}^\prime \sum_{PQ} w_l \left[ \sum_{ia} (B_{i \vk_i a \vk_a}^P)^* B_{i \vk_i a \vk_a}^Q e^{-(\varepsilon_{a\vk_a}- \varepsilon_{i\vk_i}) t_l} \right]
\left[ \sum_{jb} (B_{j \vk_j b \vk_b}^P)^* B_{j \vk_j b \vk_b}^Q e^{-(\varepsilon_{b\vk_b}- \varepsilon_{j\vk_j}) t_l} \right] \\ \nonumber
&= - \frac{1}{N_k^3} \sum_l \sum_{\vk_i\vk_a\vk_j\vk_b}^\prime \sum_{PQ} w_l M_{k_i k_a}^{PQ}(t_l) M_{k_j k_b}^{PQ}(t_l)
\end{align}
where
\begin{align}
\label{M_tensor}
    M_{k_i k_a}^{PQ}(t_l) = \sum_{ia} (B_{i \vk_i a \vk_a}^P)^* B_{i \vk_i a \vk_a}^Q e^{-(\varepsilon_{a\vk_a}- \varepsilon_{i\vk_i}) t_l} .
\end{align}
The implementation can be seen in the pseudo-code algorithm in Alg.\ref{alg:cap}. To retrieve molecular orbital (MO) energy pairs $\varepsilon_{a\vk_a}- \varepsilon_{i\vk_i}$, we start with a HF SCF calculation. 
Using this, we will determine the parameters $E_{min}$, $E_{max}$, and $R$ needed to obtain the quadrature points for Laplace integration.
Then we will calculate the RI ERI tensor $ B_{i \vk_i a \vk_a}^P$ in Eq.\ref{eq:ri_tensor}, and subsequently the $M_{k_i k_a}^{PQ}(t_l)$ tensor in Eq.\ref{M_tensor}. Lastly, we will loop over $\vk$-points using the conservation of crystal momentum, and calculate $E^{\mathrm{corr}}_{\mathrm{os}}$ (Eq. \ref{eq:E_os_RILT}.

The highest scaling step in this SOS-RILT-MP2 algorithm is calculating the $M_{k_i k_a}^{PQ}(t_l)$ tensor in Eq.\ref{M_tensor}. This step has a scaling of $N_k^2N_lN_{aux}^2N_oN_v$, where $N_k$ represents the number of $\vk$-points sampled in the Brillouin zone, $N_l$ the number of quadrature points of the LT, $N_{aux}$ the number of auxiliary basis functions, $N_o$ the number of occupied MOs, and $N_v$ the number of virtual MOs. In contrast, the highest scaling step in RI-MP2 is the calculation of the ERIs using RI shown in Eq.\ref{2eri_ri}, and has a scaling of $N_k^3N_o^2N_v^2N_{aux}$. This comparison shows that the use of the SOS-RILT-MP2 algorithm reduces the formal scaling with the number of atoms in the unit cell from $N^5$ in RI-MP2 to $N^4$, as the number of quadrature points $N_l$ to obtain a $\mu$H accuracy is small (see results below). It also has a reduced scaling with the number of $\vk$-points from $N_k^3$ to $N_k^2$. This is a great advantage for modeling solids with complex unit cells, such as point defects which usually require a large unit cell in order to reduce the periodic artifact interactions from other defects in neighboring cells. In the following section, we present the results that compare SOS-RILT-MP2 with conventional MP2 (i.e., RI-MP2).

\begin{algorithm}
\caption{Pseudo code for calculating SOS-RILT MP2}\label{alg:cap}
\begin{algorithmic}
    \State Collect MO energy pairs $\varepsilon_{a\vk_a}- \varepsilon_{i\vk_i}$ from SCF calculations
    \State Calculate $E_{min}, E_{max}, R$
    \State Using $R$ for retrieving the grid points and weights $t_l,w_l$ for the LT quadrature 
    \State Calculate the RI-ERI tensor $ B_{i \vk_i a \vk_a}^P$
    \State Calculate $M_{k_i k_a}^{PQ}(t_l)$
\Loop\ over {$k_i, k_j, k_a$}
     \State Calculate $k_b$ from conservation of crystal momentum
     \State $E_{os} \gets E_{os} = - \frac{1}{N_k^3} \sum_l \sum_{PQ} w_l M_{k_i k_a}^{PQ}(t_l) M_{k_j k_b}^{PQ}(t_l)$  
\EndLoop
\end{algorithmic}
\end{algorithm}

\section{Computational Details}
\label{computational}
To evaluate the performance of the SOS-RILT-MP2 algorithm, we implemented it in PySCF \cite{sun2018pyscf,Sun2020}. To achieve scalability benefits over conventional MP2, we applied the algorithm to both molecular and solid systems. The molecular systems included a series of linear alkane chains of increasing length: C$_{10}$H$_{22}$, C$_{20}$H$_{42}$, C$_{30}$H$_{62}$, C$_{40}$H$_{82}$, and C$_{50}$H$_{102}$. The solid-state systems included two insulators: diamond (C) and
aluminum nitride (AlN). 
These choices allowed us to rigorously assess the algorithm's performance in terms of accuracy and cost, examining the influence of increasing number of atoms within the unit cells in the molecular alkane chain calculations, and the impact of an escalating number of k-points and virtual orbitals in the solid system calculations. 
Additionally, we tested a benzene molecular crystal to evaluate the algorithm’s performance on a complex system with many atoms per unit cell and a large number of $k$-points, presenting a more challenging case for system size. 
Periodic boundary conditions were adopted for all the systems under study.
For the alkane molecules, we performed the calculations at the $\Gamma$ point using a large unit cell with a vacuum region of approximately 30 \AA\  in all three dimensions, preventing interactions between neighboring molecules to ensure isolated conditions. For the solid systems, including diamond, AlN, and benzene crystal, we performed the calculations using k-point grids with a uniform Monkhorst-Pack mesh ranging from $N_k=1\times1\times1$ to $5\times5\times5$ in the Brillouin zone that includes the $\Gamma$ point. To mitigate the issue of the divergent exchange term in periodic Hartree-Fock (HF) calculations, we used a Madelung constant correction scheme~\cite{paier2005perdew, broqvist2009hybrid, sundararaman2013regularization}. This correction ensures that both the total energies and the orbital energies converge toward the thermodynamic limit (TDL) as $N_k^{-1}$, with examples presented in our recent paper ~\cite{goldzak2022accurate}. 

Furthermore, we conducted two distinct sets of calculations: all-electron calculations and pseudopotential calculations. In the pseudopotential calculations, we tested the molecular alkane chains and the three solid systems (diamond, AlN, and benzene crystal). In these calculations, we replaced the core electrons with Goedecker-Teter-Hutter (GTH) pseudopotentials~\cite{goedecker1996,hartwigsen1998} and used the GTH-cc-pV$X$Z correlation-consistent Gaussian basis sets that are optimized for periodic calculations with GTH pseudopotentials~\cite{Ye22JCTC}. Specifically, the double-zeta GTH-cc-pVDZ was used for the molecular chains given the large number of atoms, up to triple-zeta GTH-cc-pVXZ (X=D,T) were used for benzene crystal to manage memory constraints, and up to quadruple-zeta GTH-cc-pVXZ (X=D,T,Q) were tested for diamond and AlN.

In the all-electron calculations, we tested the diamond solid system and the molecular chains, excluding C$_{50}$H$_{102}$ due to computational constraints. 
In these calculations, we used Dunning's original cc-pVXZ basis set \cite{dunning1989gaussian}. Again, the double-zeta cc-pVDZ was used for the molecular chains, and up to quadruple-zeta cc-pVXZ (X=D,T,Q) were tested for diamond.
For the calculation of the ERIs in all basis sets, we used the Gaussian density fitting (GDF) method~\cite{Sun2017a} with the cc-pVDZ-RI auxiliary basis for the molecules and benzene crystal and cc-pVQZ-RI for diamond and AlN.

All Calculations were preformed on Intel Xeon Gold 6338 205W processors $@$ 2.0 GHz with 4 cores.

\section{Results}
\label{results}
\begin{figure}[h!]
\centering
\includegraphics[width=0.7\textwidth]{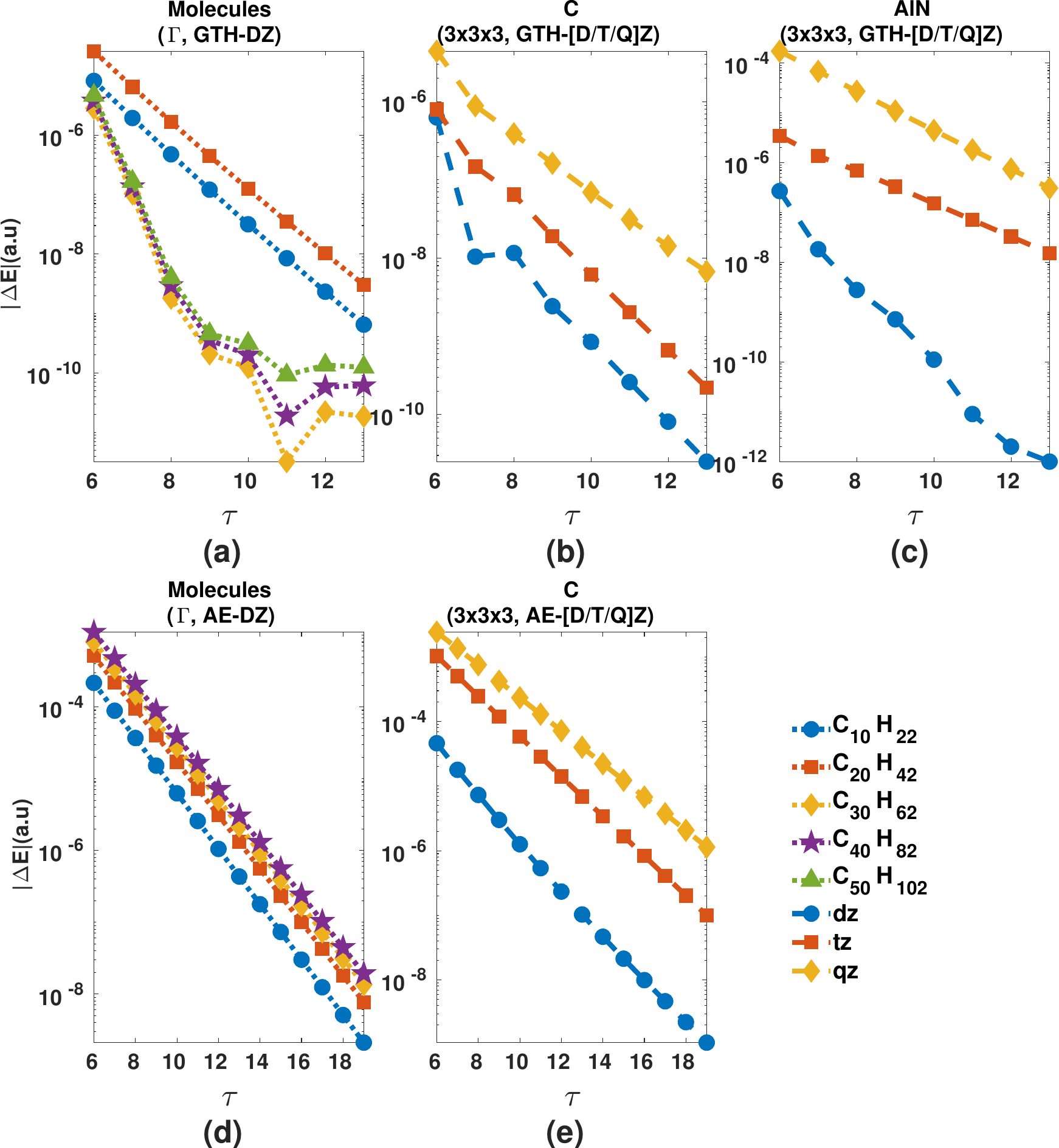}
\caption{Log scaled absolute energy difference between the opposite spin correlation energies (a.u) of the SOS-RILT MP2 algorithm and the conventional MP2 as a function of different quadrature points. (a) Linear alkane chains with increasing lengths (C$_{10}$H$_{22}$ to C$_{50}$H$_{102}$) calculated using the GTH pseudopotential and the GTH-cc-pVDZ basis set at the Gamma point. (b) Diamond calculated with the GTH pseudopotential and the GTH-cc-pVXZ (X=D,T,Q) basis sets using a 3x3x3 k-point grid. (c) AlN calculated with the GTH pseudopotential and the GTH-cc-pVXZ (X=D,T,Q) basis sets using a 3x3x3 k-point grid. (d) Linear alkane chains with increasing lengths (C$_{10}$H$_{22}$ to C$_{40}$H$_{82}$)  calculated with the all-electron cc-pVDZ basis set at the Gamma point. (e) Diamond calculated with the all-electron cc-pVXZ (X=D,T,Q) basis sets using a 3x3x3 k-point grid.}
\label{fig:sos_energy}
\end{figure}

As a first step, to verify the accuracy of the LT part in the SOS-RILT-MP2 algorithm, we compared the opposite-spin energy term, $E^{corr}_{os}$, in the SOS-RILT-MP2 algorithm (Eq.~\ref{eq:mp2_os}) to that in the conventional MP2 algorithm. Since both SOS-RILT-MP2 and conventional MP2 use RI, the difference between them can only be due to the LT component. Specifically, we examined the absolute energy difference between these terms at different quadrature points customized to each calculation set. For the set involving all-electron calculations, we expanded the range of quadrature points from 6 to 19. In contrast, for the set using GTH pseudopotentials, we restricted our analysis to quadrature points between 6 and 13. Note that these evaluations for the solid systems were performed using a 3x3x3 k-point grid. The findings, depicted in Fig.~\ref{fig:sos_energy}, reveal a consistent pattern in the tested systems.

Across all panels, we observe that increasing the number of quadrature points exponentially reduces the LT error (i.e., the energy difference between conventional MP2 and SOS-RILT-MP2) in every system, as expected. In other words, with more quadrature points, the SOS-RILT-MP2 method steadily improves in precision. Practically, this improvement is so rapid that only a relatively small number of quadrature points is required to achieve a $\mu$H accuracy. This trend holds for alkane chains, diamond, and AlN. A similar convergence behavior was also observed for benzene crystal, though not shown here. These findings highlight the robustness and consistency of the SOS-RILT-MP2 algorithm across a wide range of chemical structures. 

Examining the effect of the basis set size on the energy difference in the solids (panels (b), (c), and (e)), we see that a larger basis set amplifies the LT error in both all-electron and pseudopotential calculations. The reason is straightforward: a larger basis set, applied to the same system, includes more orbitals and thus increases the maximum orbital energy across all k-points, \(\varepsilon_{{\max}\vk_{max}}\). As a result, \(E_{\max}\) and \(R\) both increase, making a higher quadrature point necessary to maintain the same accuracy (see Eqs.~\ref{num_int}-\ref{num_int2}).

A similar size effect is evident for the alkene chains. In the all-electron calculations shown in panel (d), increasing the system size, i.e., the chain length, further widens the LT error. This occurs because a larger system increases \(\varepsilon_{{\max}\vk_{max}} - \varepsilon_{{\min}\vk_{\min}}\), and consequently \(E_{\max}\) and \(R\), leading to a higher number of quadrature points required to preserve the same level of accuracy.

However, in the pseudopotential calculations in panel (a), there is a different behavior. There, we can distinguish between two subsets: C$_{10}$H$_{22}$ and C$_{20}$H$_{42}$ versus C$_{30}$H$_{62}$, C$_{40}$H$_{82}$, and C$_{50}$H$_{102}$. Within each subset, the trend holds: as system size increases, the LT error increases as well. Yet a discrepancy appears when moving from C$_{20}$H$_{42}$ to C$_{30}$H$_{62}$, where the difference decreases instead of increasing. This sharp discontinuity arises from a numerical artifact in the calculations. Yet, even with this artifact, these findings demonstrate that the SOS-RILT-MP2 method achieves highly accurate results at the $\mu H$ level, with a small number of quadrature points.

Furthermore, looking at Fig.~\ref{fig:sos_energy}, we can notice an additional trend. Comparing panels (a) and (b) with panels (c) and (d) reveals that the all-electron calculations demand more quadrature points to reach the same level of LT error as in the pseudopotential calculations. For instance, in the alkene chains, panel (a) shows that a quadrature point of 8 is suffices for an acceptable $\mu H$ difference between the methods in the pseudopotential calculations, while panel (d) indicates a higher quadrature point of 14 is needed for the all-electron calculations. This difference in optimal quadrature points can be attributed to the different orbital energies obtained in each type of calculation. In the all-electron case, we calculate the core orbital energies, while in the pseudopotential case we do not. Therefore, in the all-electron calculations, the core orbital energies drive down \(\varepsilon_{{\min}\vk_{\min}}\), and hence push up \(E_{\max}\) and \(R\). As a result, more quadrature points are needed to achieve a similar level of accuracy compared to the pseudopotential case.

Overall, these findings confirm the reliability and high accuracy at the $\mu H$ level with a small number of quadrature points for the SOS-RILT-MP2 method, while highlighting its potential for broader application in complex computational chemistry scenarios. In particular, they highlight the method’s excellent performance in both all-electron and pseudopotential calculations, which is especially relevant for problems involving heavy elements and core excitations. A recent study has shown that all-electron calculations significantly improve accuracy in properties such as the bulk modulus of heavy-element solids, capturing crucial relativistic effects.~\cite{gaurav2024challenges}
Moreover, $G_0W_0$ studies based on Gaussian basis sets have further demonstrated the value of all-electron methods, accurately determining quasiparticle energies, band structures, and both valence and core excitations in weakly correlated materials.~\cite{zhu2021all} Taken together, these advances stress the importance of a robust method, like SOS-RILT-MP2, that can handle both all-electron and pseudopotential calculations, to enable a more comprehensive understanding of complex electronic systems.

\begin{figure}[h!]
\centering
\includegraphics[width=0.7\textwidth]{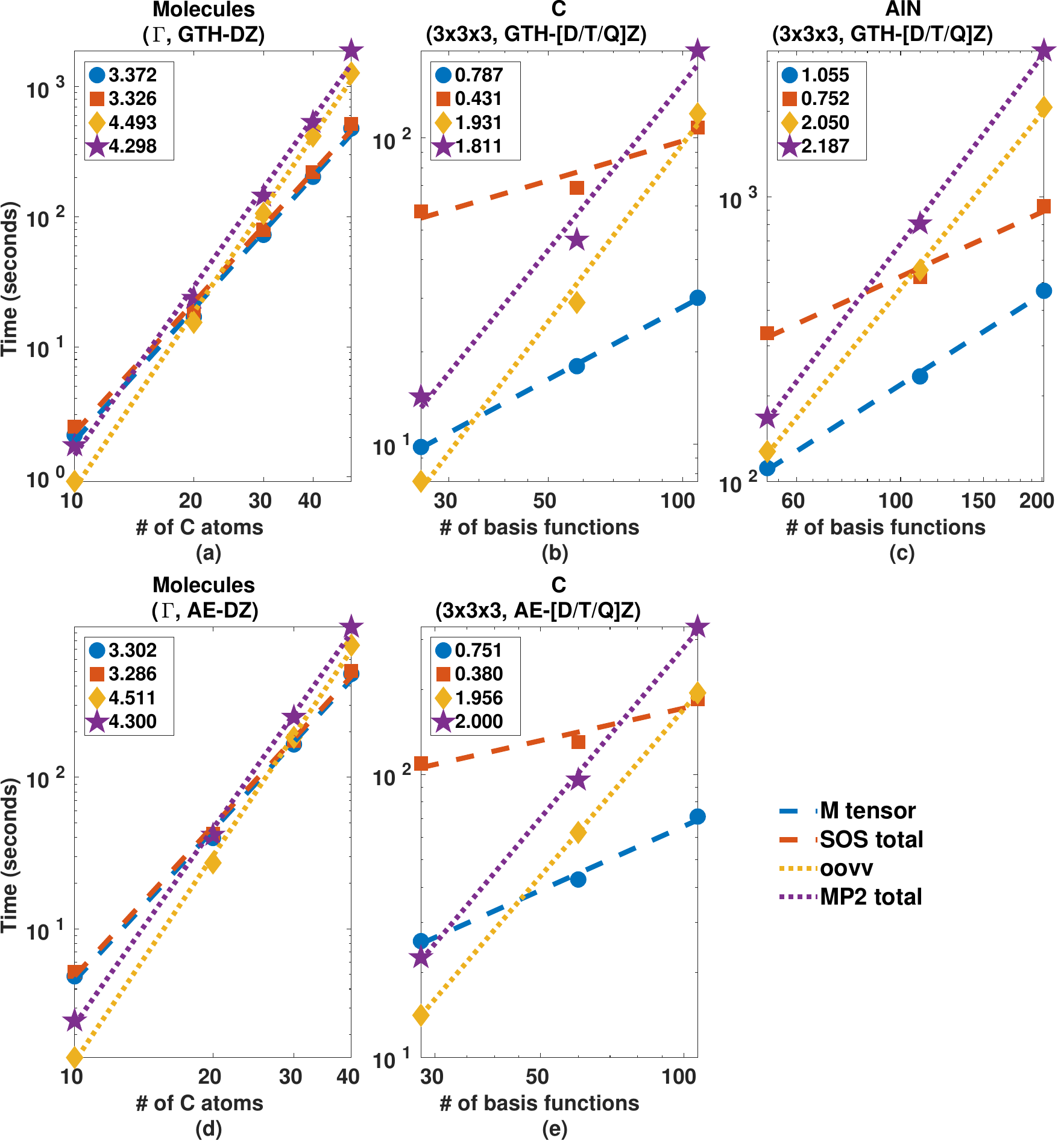}
\caption{Log scaled time consumption (in seconds) for the $\bold{M}$ tensor step of the SOS-RILT MP2 algorithm ("$\bold{M}$ tensor") , whole of the SOS-RILT MP2 algorithm ("SOS total"), the oovv step of the conventional MP2 algorithm ("oovv"), and  whole of the conventional MP2 algorithms ("MP2 total"), as function of  different system complexities. (a) as a function of the number of carbon atoms in linear alkane chains of increasing length (C$_{10}$H$_{22}$ to C$_{50}$H$_{102}$) calculated using a gth-pseudopotential and an optimized cc-PVDZ basis set at the Gamma point (b) as a function of the number of optimised basis functions (cc-pVDZ to cc-pVQZ) used calculating Diamond with a gth-pseudopotential using a 3x3x3 k-point grid (C) as a function of the number of optimised basis functions (cc-pVDZ to cc-pVQZ) used calculating AlN with a gth-pseudopotential using a 3x3x3 k-point grid (d) as a function of the number of carbon atoms in linear alkane chains of increasing length (C$_{10}$H$_{22}$ to C$_{40}$H$_{82}$) calculated using an all-electron approach and a cc-PVDZ basis set at the Gamma point (e) as a function of the number of basis functions (cc-pVDZ to cc-pVQZ) used calculating Diamond with an all-electron approach using a 3x3x3 k-point grid. For each data set, trend lines are presented, with the legend indicating their slopes.}
\label{fig:sos_time}
\end{figure}

In the second part of our analysis, we aimed to confirm the scaling of the most time-consuming component in each algorithm: that is, constructing the \({M_{k_i k_a}^{PQ}(t_l)}\) tensor (referred to as the “\(\bold{M}\) tensor”, see Eq.~\ref{M_tensor} and Alg.~\ref{alg:cap}) for SOS-RILT-MP2, and calculating the ERIs with the auxiliary basis (referred to as the ``oovv step'', see Eq.\ref{2eri_ri}) for conventional MP2. 

We examined the time consumption for each step and the total time consumed by each algorithm as a function of the system's complexity: For the alkane chains, we focused on scaling as the number of atoms increased. In solid systems, we explored the impact of the basis set size on the computation time, specifically using a 3x3x3 k-point grid. It should be noted that, based on Fig.~\ref{fig:sos_energy}, to achieve an acceptable $\mu$H difference in energy between the two algorithms for all of the systems in question, we employed nineteen quadrature points for the all-electron calculations and eleven quadrature points for the pseudopotential calculations. Note that for each solid, the number of functions in the auxiliary basis set and the number of occupied orbitals remained constant across the different basis sets, while only the number of virtual orbitals increased as  the basis set grew larger. 
In this case, because the oovv step scales as \(N_k^3 N_o^2 N_v^2 N_{aux}\) (see Eq.~\ref{2eri_ri}), its time consumption is expected to scale with an exponent of 2 relative to the basis set size, since only the number of virtual orbitals ($N_v$) changes. By contrast, for the \(\bold{M}\) tensor, which scales as \(N_k^2 N_l N_{aux}^2 N_o N_v\) (see Eq.~\ref{M_tensor}), the time consumption is expected to scale with an exponent of 1 with respect to the basis set size. 
However, in the molecular case, the number of functions in the auxiliary basis set and the number of occupied and virtual molecular orbitals all increase as the molecular chain length grows. Consequently, the time consumption for the \(\text{oovv}\) step is expected to scale with an exponent of 5 relative to the chain size, whereas the time consumption for the \(\boldsymbol{M}\) tensor is anticipated to scale with an exponent of 4 with respect to the basis set size.

The results for this part of our analysis (see Fig.\ref{fig:sos_time}) confirm the theoretical predictions regarding the scaling of the most time-intensive components in both algorithms, i.e. the \(\bold{M}\) tensor and the oovv step.  In all of our systems (panels (a) to (e) in Fig.\ref{fig:sos_time}), the \(\bold{M}\) tensor scales at least one unit less than the oovv step. A similar trend is also apparent for the total time consumed by each algorithm, where the total time consumed by the SOS-RILT-MP2 algorithm scales at least one unit less than the total time consumed by the MP2 algorithm. One of the biggest differences between the algorithms can be seen in panel (e) in Fig.\ref{fig:sos_time}, showing the time scaling for an all-electron diamond calculation. In this panel, we can see that both the oovv step and the total time for conventional MP2 scale significantly higher than the \(\bold{M}\) tensor and the total time for SOS-RILT MP2, with a slope of 1.956 and 2.000 vs. 0.751 and 0.380 respectively. These results demonstrate the superior efficiency and scalability of the SOS-RILT-MP2 algorithm in calculating complex unit cells with a large set of bases in periodic systems. 

\begin{figure}[h!]
\centering
\includegraphics[width=0.4\textwidth]{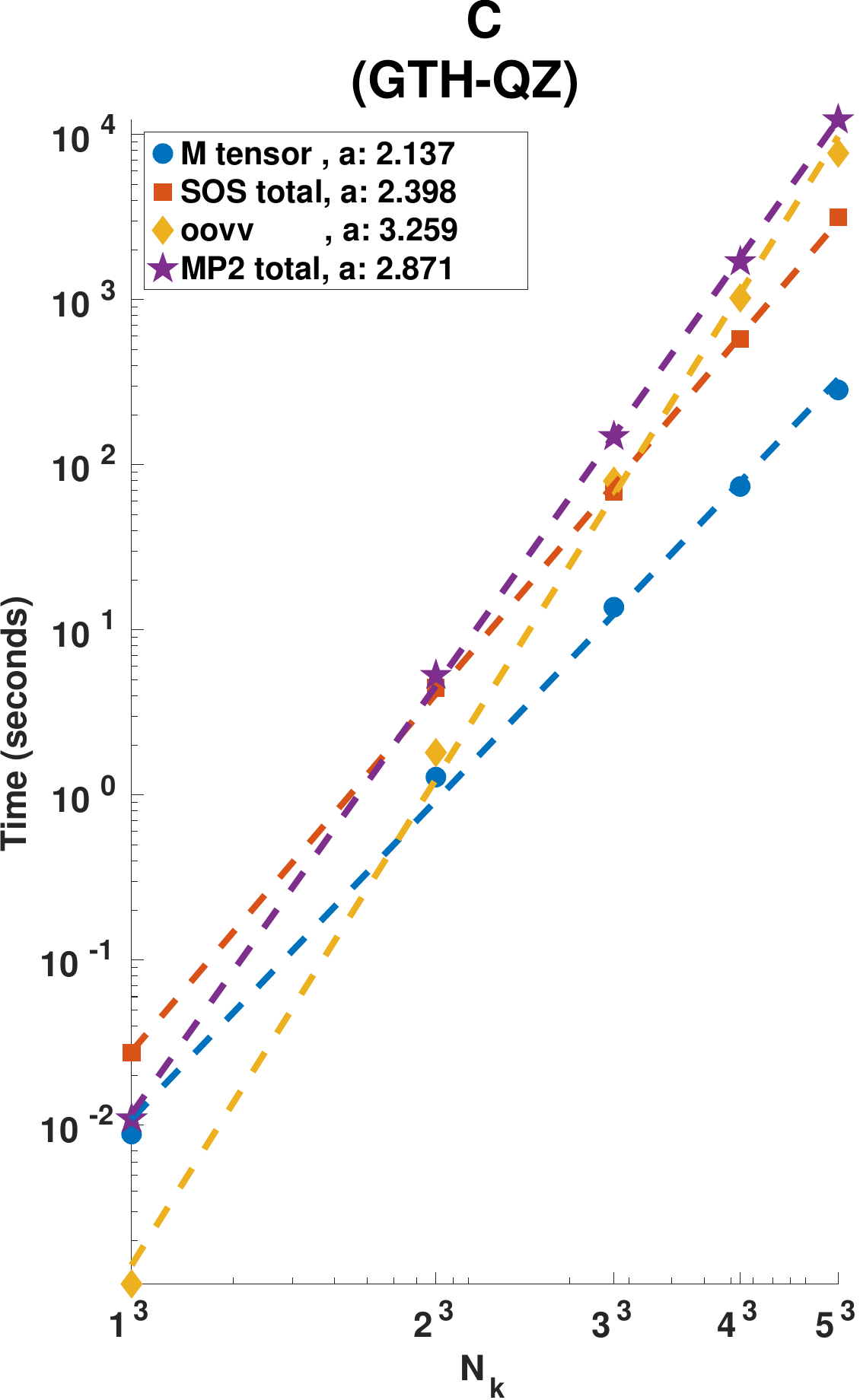}
\caption{Log scaled time consumption (in seconds) for the $\bold{M}$ tensor step of the SOS-RILT MP2 algorithm ("$\bold{M}$ tensor") , whole of the SOS-RILT MP2 algorithm ("SOS total"), the oovv step of the conventional MP2 algorithm ("oovv"), and  whole of the conventional MP2 algorithms ("MP2 total"), as function of the log scaled number of grid k-points ($N_k$) (1x1x1 to 5x5x5) for Diamond using optimized cc-pVQZ basis set. Each data set includes a trend line, labeled in the legend as "Dataset Name, a = [slope value]".}
\label{fig:sos_kpnts}
\end{figure}

To reinforce our conclusion on the efficiency of the SOS-RILT-MP2 algorithm, we tested it on the benzene molecular crystal. Due to memory constraints, we used a 2$\times$2$\times$2 k-point grid with GTH-cc-pVDZ and GTH-cc-pVTZ basis sets, and a 3$\times$3$\times$3 k-point grid with GTH-cc-pVDZ basis set~\cite{Ye22JCTC}. The results in Table~\ref{tab:benzene_timing} clearly demonstrate SOS-RILT-MP2’s superior scalability. For example, in the 3$\times$3$\times$3 k-point grid case, the oovv step in conventional MP2 took about 276 hours, and the total time 364 hours. On the contrary, for the same case, the \(\mathbf{M}\) tensor step in SOS-RILT-MP2 took only 38 hours, and the total time 47 hours. These results highlight the significantly improved efficiency of SOS-RILT-MP2, making it particularly effective for periodic systems, and especially those with large and complex unit cells.

\begin{table}[h]
    \centering
    \begin{tabular}{c|cccc}
        k-point grid \&  cc-pV(XZ) basis set & \(\mathbf{M}\) tensor & SOS-RILT-MP2 total & oovv & MP2 total \\
        \hline
        2x2x2 \& cc-pVDZ & 0.996 & 1.113 & 6.732 & 8.190 \\
        2x2x2 \& cc-pVTZ & 3.559 & 8.077 & 49.176 & 66.143 \\
        3x3x3 \& cc-pVDZ & 37.568 & 46.975 & 275.772 & 363.655 \\
    \end{tabular}
    \caption{Comparison of computation times (hours) for the \(\mathbf{M}\) tensor and the oovv steps, along with total times for SOS-RILT-MP2 and conventional MP2 algorithms across different basis sets and k-point grids for the Benzene molecular crystal. Quadrature points for SOS-RILT-MP2 were chosen to ensure $\mu$Hartree precision, with 12 points for DZ and 15 for TZ.}
    \label{tab:benzene_timing}
\end{table}

In the last step of our analysis, which focused solely on the pseudopotential calculations, we explored how both algorithms' time-intensive components scale with the number of k-points sampled in the Brillouin zone. We specifically analyzed the algorithm's performance for diamond using the GTH-cc-pVQZ basis set and k-point grids ranging from $N_k=1\times1\times1$ to $N_k=5\times5\times5$. Our analysis of the \(\bold{M}\) tensor and the oovv step, along with the total time consumption for each algorithm, is presented in Fig.~\ref{fig:sos_kpnts}. The results indicate that the \(\bold{M}\) tensor and oovv scale roughly close to theoretical values (2.137 and 3.259 respectively, compared to theoretical expectations of 2 and 3). Although the scaling of the total time for conventional MP2 was slightly smaller than that of the oovv step (2.871 vs. 3.259), the scaling of the total time for SOS-RILT-MP2 was slightly larger than that of the \(\bold{M}\) tensor (2.398 vs. 2.137). This discrepancy can be attributed to the scaling of the \(\bold{M}\) tensor step as \(N_k^2\), while the final part of the algorithm scales as \(N_k^3\) (see Alg.~~\ref{alg:cap}). However, it is worth mentioning that even with this discrepancy, comparing the total time for the conventional MP2 algorithm in the 5$\times$5$\times$5 case with that of the SOS-RILT-MP2 algorithm, we can see a clear difference of about one order of magnitude in favor of SOS-RILT-MP2. 
This analysis further demonstrates the efficiency and scalability advantages of SOS-RILT-MP2 over conventional MP2, showing superior performance in handling larger k-point grids, regardless of the number of virtual orbitals. This is a huge advantage for reaching the thermodynamic limit and mitigating basis-set errors, as well as for calculating complex unit cell systems that require large k-point meshes. 

\section{Conclusions}
\label{conc}

In this work, we present a novel SOS-RILT-MP2 algorithm for periodic systems, which reduce the scaling of the most time-consuming step in conventional MP2 from $N_k^3 N^5$ to $N_k^2 N^4$, by using the resolution of the identity approximation combined with the Laplace transform algorithm. We outline the methodology and algorithmic steps of this approach, utilizing periodic Gaussian basis sets. These basis sets enable calculations to be performed both with and without pseudopotentials.

We tested the algorithm both on the molecular systems with increasing the number of atoms and on solids with increasing the basis set size and the number of k-points. In both cases, we showed that the Laplace transform part of the algorithm can reach micro-Hartree precision with a small number of quadrature points; the maximal was 19 for diamond with an all-electron, quadruple-zeta basis set. We verified the efficiency of the algorithm by testing the scaling of the algorithm with increasing the basis set size and the number of k-points. We show that the SOS-RILT-MP2 algorithm has reduced the scaling of conventional MP2 by an order of magnitude with the number of atoms (or basis functions) in the unit cell, as well as reduced scaling with the number of k-points. We also tested our efficient algorithm on the benzene molecular crystal, achieving a $\sim$8$\times$ speedup and a saving of 317 hours compared to conventional MP2s. This significantly aids in achieving the thermodynamic limit and reducing basis-set errors, as well as in calculating complex unit cell systems that necessitate extensive k-point meshes.  

Previous studies on periodic MP2 for molecular crystals observed good performance for SOS-MP2 in cohesive energy calculations for some crystals ~\cite{del2012second} and an underestimation for others such as benzene ~\cite{bintrim2022integral}. In our last work, we applied SOS-MP2 on 12 semiconductors and insulators and calculated the lattice constant, bulk modulus, and cohesive energies and showed promising results in various scaling parameters. Future work should focus on applications of the adsorption of molecules on surfaces for catalysis, defects in materials, and other complex interfaces. A recent paper calculating the adsorption of the CO molecule on the MgO surface showed that MP2 yields good results for the adsorption energy and vibrational frequencies \cite{ye2024adsorption}. 
Furthermore, extension of the periodic SOS-RILT algorithm to excited state methods (such as CIS(D) and CC2) should give great advantage in calculating the optical properties of these materials.

\section*{Data Availability}
The input and output data files associated with this study and all analysis are available from the corresponding author, T.G, upon reasonable request. 
The source code for periodic SOS-RILT-MP2 is available from the corresponding author, T.G, upon reasonable request.

\begin{acknowledgement}
We thank Timothy Berkelbach for helpful discussions. X.W. acknowledges the start-up funding from the University of California, Santa Cruz. T.G acknowledges funding from the Ministry of Innovation, Science and Technology Israel grant No.5802, and the Ministry of Energy Israel.
\end{acknowledgement}

\bibliography{mp21,ISFbib,sos_LT}
\end{document}